\documentclass[]{article}
\usepackage[top=1in, bottom=1in, left=1.25in, right=1.25in]{geometry}
\usepackage{graphicx}
\usepackage{authblk}

\title{A compact, thermal noise limited reference cavity for ultra-low noise microwave generation}

\author[1,*]{J. Davila-Rodriguez}
\author[1,2]{F.N. Baynes}
\author[1]{A.D. Ludlow}
\author[1]{T. M. Fortier}
\author[1,3]{H. Leopardi}
\author[1,3]{S.A. Diddams}
\author[1,$\dagger$]{F. Quinlan}

\affil[1]{NIST, Time and Frequency Division, 325 Broadway MS 847, Boulder, CO 80305, USA}
\affil[2]{Current Address: Institute for Photonics and Advanced Sensing and School of Physical Sciences, University of Adelaide, Adelaide, SA 5005, Australia}
\affil[3]{ Department of Physics, University of Colorado Boulder, 440 UCB Boulder, Colorado, 80309, USA}

\affil[*]{Corresponding author: josuedavila@gmail.com}
\affil[$\dagger$]{franklyn.quinlan@nist.gov}

\begin{document}

\maketitle

\begin{abstract}
We demonstrate an easy to manufacture, 25 mm long ultra-stable optical reference cavity for transportable photonic microwave generation systems. Employing a rigid holding geometry that is first-order insensitive to the squeezing force and a cavity geometry that improves the thermal noise limit at room temperature, we observe a laser phase noise that is nearly thermal noise limited for three frequency decades (1~Hz to 1~kHz offset) and supports 10 GHz generation with phase noise near \mbox{-100 dBc/Hz} at 1 Hz offset and \mbox{$<$-173 dBc/Hz} for all offsets \mbox{$>600$ Hz}. The fractional frequency stability reaches $2\times10^{-15}$ at 0.1 s of averaging.
\end{abstract}

\section{Introduction}

Continuous wave lasers locked to ultra-stable cavities deliver extraordinarily pure electromagnetic waves, reaching a frequency stability of 10$^{-16}$ at 1 s \cite{Hafner15,Kessler2012}. These signals have therefore served as a tool in experimental physics from precision spectroscopy \cite{JanisAlnisPRA2008} and optical atomic frequency standards \cite{Schioppo2016}, to gravitational wave detection \cite{LIGO_Cavities} and tests of fundamental physics \cite{IsotropyTest}. The utility of ultra-stable lasers can be extended to the rf and microwave domain via optical frequency division (OFD) \cite{Fortier2011}, where a femtosecond optical frequency comb is phase locked to the stable optical frequency reference. This coherent division of an optical signal to the microwave domain results in phase noise power \mbox{$\sim$90 dB} lower than that of the optical reference, yielding some of the lowest phase noise microwave signals produced by any means \cite{Fortier2011,syrte2016}. Such low noise microwaves have the potential to contribute in several applied and fundamental areas such as radar \cite{Scheer1993}, transduction of quantum states between microwave and optical fields \cite{AndrewsLehnert2014}, and improving the performance of microwave atomic frequency standards such as cesium fountain clocks \cite{Santarelli1999}.

The frequency stability and phase noise of a cavity-stabilized laser is ultimately limited by the length stability of the reference cavity \cite{NumataPRL,harry2012optical}. Thermally driven fluctuations, primarily in the mirrors and coatings, set a fundamental limit to the cavity length stability, the impact of which is reduced in state-of-the-art systems by extending the cavity length \cite{Young1999}, or by operating at cryogenic temperatures \cite{Hafner15}. However, for many applications, including those in the microwave domain, it is desirable to have a stable laser that is compact, rigidly-held, vibrationally insensitive and mobile, thereby allowing operation outside the staid laboratory environment. Additionally, in contrast to optical clock applications, many microwave applications require low noise performance in the millisecond to microsecond regime. Given the broad phase locking bandwidth of some frequency combs used for OFD \cite{Quinlan2014}, the phase noise of the optical reference at millisecond time scales can directly impact the microwave phase noise. Demonstration of low noise performance of the cavity stabilized laser out to \mbox{$\sim$1 MHz} offset frequency is therefore critical.

In this letter we propose and demonstrate a rigidly held cavity with an easily manufacturable cylindrical design only 25 mm in length. For offset frequencies from 1~Hz to 1~kHz, near thermal-noise-limited performance is demonstrated, translating to a 10~GHz microwave with 1~Hz phase noise at \mbox{-97 dBc/Hz} and 1~kHz phase noise below \mbox{-185 dBc/Hz}. Characterization of the phase noise out to 1 MHz offset indicates the support of phase noise on a 10~GHz carrier below \mbox{-173 dBc/Hz} for offset frequencies beyond 600~Hz. These results address the practical challenges of having a simple, transportable cavity for microwave applications while simultaneously providing low phase noise.

\section{Cavity Design}
Previous designs of centimeter scale, rigidly-held reference cavities \cite{Leibrandt2011Spherical,LeibrandtPRA2013,Webster2011,DidierTriangle2016,
Parker2014,Dariusz2016} have predicted or demonstrated extraordinarily low levels of vibration sensitivity, either passively or after feed-forward correction from inertial sensors \cite{LeibrandtPRA2013}. An important task in the case of rigidly-held cavities is finding a geometry which is minimally sensitive to both the holding force and the vibrations coupled through the holding structure. To date, cavity geometries include a spherical spacer \cite{Leibrandt2011Spherical}, a cubic spacer with truncated corners held using a tetrahedral symmetry \cite{Webster2011}, and a triangular cavity \cite{DidierTriangle2016}.

For our cavity, shown in Fig. 1, we have chosen a simple cylindrical spacer with a large radius-to-length ratio. This cavity geometry allows for the existence of a holding radius where the cavity can be squeezed without affecting its length to first order. This effect can be understood by comparing the expected behavior from squeezing a cylinder with finite elasticity (Poisson`s ratio $>$ 0) on its axis and along the rim. It would be expected that the cylinder's axis will compress in the former case and bulge in the latter. The squeeze insensitive point is the diameter at which these two effects cancel. We have performed finite-element analysis to verify our intuition and find the location of this point, the results of which are shown in Fig. 1(c). For a 25~mm long spacer, we find that the zero crossing of the holding force sensitivity exists for spacer diameters larger than \mbox{$\sim$40 mm}. We have chosen a diameter of 50 mm as a compromise between the location of the squeeze insensitive point being reasonably removed from the edge of the spacer and keeping the spacer’s volume constrained.

\begin{figure}[htbp]
\centering
\includegraphics[width=3.3 in]{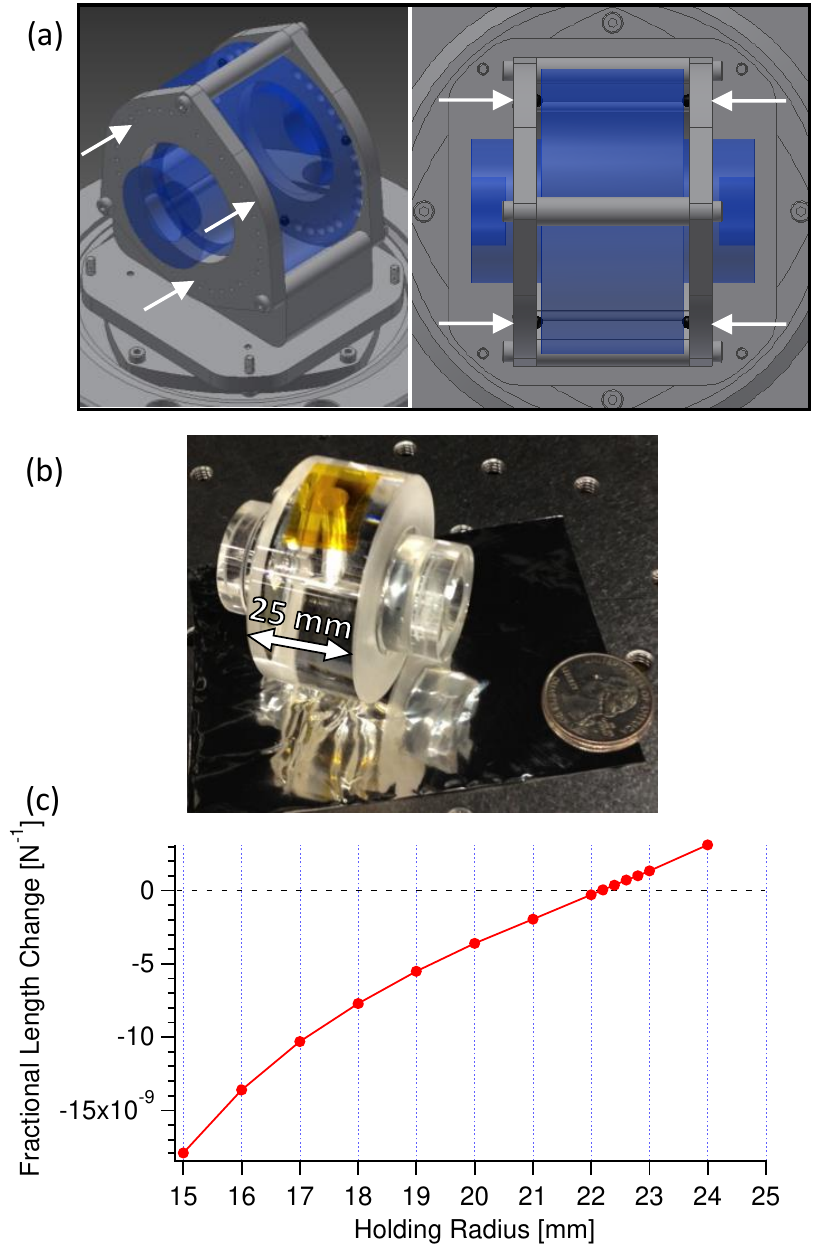}
\caption{(a) Drawing of the cavity and Invar cavity holder. The arrows indicate the location of the elastomer balls used as point contacts to rigidly hold the cavity. An additional radiation shield (not shown) covers the cavity holder assembly. (b) Photograph of the cavity. (c) The calculated cavity fractional length change as a function of the holding radius. Note the zero crossing near 22 mm.}
\label{fig:false-color}
\end{figure}

The cavity spacer is made out of Corning ultra-low expansion (ULE) \cite{NISTDisclaimer} glass with a 5 mm diameter axial bore for the optical mode and an additional radial bore at the midpoint along the cylinder's length for venting the cavity. Low-loss, high-reflectivity dielectric mirrors on fused-silica substrates are optically contacted to the spacer, and additional ULE rings are contacted to the outside of each of the mirrors. The ULE rings were added to shift the zero crossing of the cavity’s coefficient of thermal expansion (CTE) to a convenient temperature \cite{Legero:10}, though for the results presented here no effort was made to hold the cavity temperature at the zero CTE point. With the current substrate and backing ring thicknesses, the entire cavity assembly is \mbox{$\sim$50 mm} long and occupies a 61~mL volume. The cavity is rigidly held in a vacuum chamber by a pair of Invar parallel plates which squeeze three 3.2 mm diameter elastomer balls on each side of the cavity. In order to easily test different holding positions, the Invar plates have been have manufactured such that the location of the holding point can be varied in \mbox{100 $\mu$m} increments around the predicted force insensitive point.

We have measured the cavity’s acceleration sensitivity at several holding positions by mounting the system on a rotatable optical breadboard and flipping all three spatial axes while monitoring the laser’s frequency. The laser remained locked to the resonance throughout the measurement. We find the largest acceleration sensitivity to be along the cavity axis at $\sim4.5\times10^{-10}$ g$^{-1}$, minimized at a holding radius of 24.5 mm. This acceleration sensitivity is larger than expected, and it may be dominated by residual asymmetries in the holding structure, or in the cavity manufacture. For subsequent characterization the cavity is mounted on an active vibration isolation platform, and the measured residual acceleration spectrum was determined not to significantly contribute to the resulting phase noise of the locked laser. 

The elimination of vibration-induced cavity length fluctuations allows for the possibility of phase noise performance at the fundamental limit, given by Brownian noise in the mirror coatings and substrates, as well as thermo-elastic, thermo-optic, and thermo-refractive noise \cite{harry2012optical}. In order to reduce the fundamental noise while maintaining a compact, room temperature design, a large optical mode is generated by choosing the mirror radius of curvature (ROC) that produces a cavity close to instability \cite{Schioppo2016}. Increasing the spot size can be achieved either by increasing the radius of curvature of the mirrors or adopting a near-concentric cavity \cite{Amairi2013}. We have chosen a plano-10.2 m ROC design, yielding an optical mode diameter \mbox{$\sim$800 $\mu$m}. This leads to a predicted thermal noise limit for our \mbox{25 mm} long cavity to be $\sim$-9 dBrad$^2$/Hz at 1 Hz offset. For comparison, a cavity with the curved mirror having a standard 50-cm ROC would need to be at least 40 mm long to obtain the same thermal noise limited phase noise. The various thermal noise contributions as well as the total thermal noise are shown in Fig. 2 (b).

\section{Results}
Using a commercially available single-longitudinal-mode fiber laser at 1070 nm, we have measured the cavity photon lifetime and calculated the finesse to be $\sim$400,000. For stabilization the laser is phase modulated using a fiber pig-tailed electro-optic modulator (EOM) and sent to the reference cavity. The reflected sidebands are demodulated to obtain a Pound-Drever-Hall (PDH) error signal \cite{Drever1983}. The laser frequency is locked with 700~kHz bandwidth by feedback to the driving frequency of an acousto-optic modulator (AOM) and to the laser cavity length for fast and slow corrections, respectively. The laser power impinging on the cavity is $\sim$70 $\mu$W and is stabilized by photodetecting a fraction of the incoming light and correcting the power driving the AOM. The setup is placed in an enclosure, but it is not actively temperature stabilized and no feedback control of the residual amplitude modulation from the EOM is applied. We have found these measures to be unnecessary, as they do not improve the phase noise for offset frequencies $>$ 1 Hz. At longer time scales both effects play a more significant role, limiting the ultimate long-term stability. However, for many applications of low-noise microwaves the stability at longer time scales is inconsequential and the reduced system complexity is advantageous. The useable output power is $\sim$3 mW.

\begin{figure}[hp]
\centering
\includegraphics[width=3.3 in]{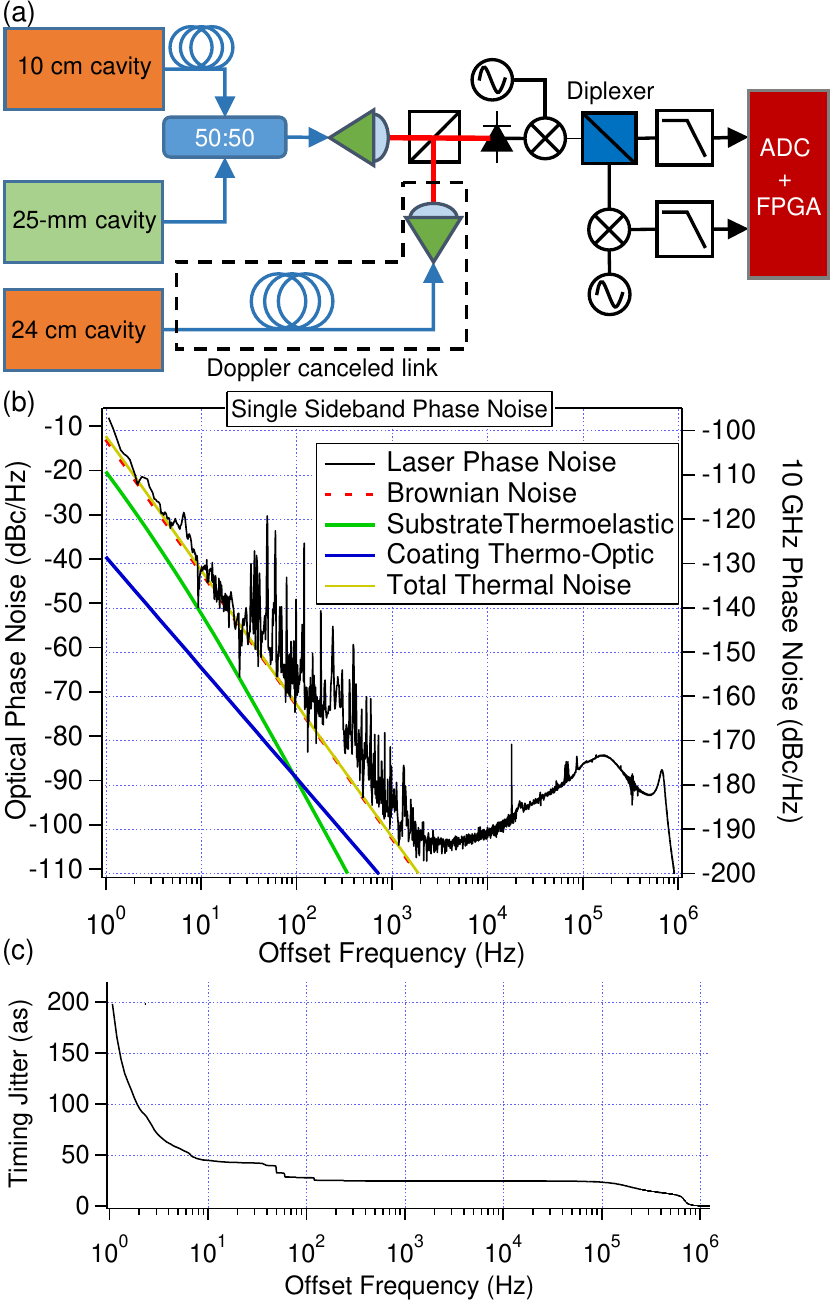}
\caption{(a) Schematic of the phase noise measurement. Beat notes with two independent references are sampled simultaneously. (b) Phase noise of the laser stabilized to the 25 mm cavity. The phase noise of our laser is recovered by averaging the cross spectrum of both beat notes. The yellow line is the total predicted thermal noise. (c) The integrated timing jitter as a function of offset frequency. The total timing jitter in the 1 Hz to 1 MHz band is $\sim$200 attoseconds.}
\label{fig:false-color}
\end{figure}

To characterize the phase noise of our cavity-stabilized laser we obtain two heterodyne beat-notes with two independent reference lasers, both of them near 1070 nm, locked to their respective cavities, as shown in Fig. 2(a). One of the references is locked to a 10 cm long cavity and has a 1 s Allan deviation of $8\times10^{-16}$. The other reference is locked to a 24 cm long cavity and has a 1 s Allan deviation of $4\times10^{-16}$. The frequencies of all three lasers are within 2 GHz of each other, allowing us to directly obtain heterodyne beats between all lasers to characterize their performance. With a combination of measurements on the individual beat-notes, we find several regions in the phase noise spectrum where the measurement is always limited by one or both of the reference lasers. To recover the phase noise of the 25 mm cavity laser, we simultaneously sampled both beat-notes \cite{Sherman2016} and subsequently calculated the cross-spectrum by averaging the complex product of the fast Fourier transform of each of the phase records. Since the noise of the reference lasers is uncorrelated, the averaging rejects their phase noise by $\sqrt{N}$ where $N$ is the number of averages.

The phase noise measurement is shown in Fig. 2 (b). Note that the laser remains nearly thermal noise limited for 3 decades (1~Hz to 1~kHz). Between 100~Hz and 1~kHz, there is a small amount of residual noise, partially due to 60 Hz harmonics from the system power sources. To achieve thermal noise limited performance, it was necessary to use at least 50 $\mu$W of power to improve the PDH sensitivity and lower the impact of the electronic noise below the thermal noise limit.  Between 700~Hz and 2~kHz, electronic noise originating within the PDH loop contributes to the phase noise, resulting in a slight increase above the thermal noise limit. Beyond 2~kHz, the phase noise of the laser increases due to limited loop gain to suppress the free-running laser noise. Also notable from the phase noise in Fig. 2 (b) is the fact that the 25 mm cavity laser supports 10 GHz generation \mbox{$<$ -190 dBc/Hz} in the 2~kHz to 10~kHz offset range, and supports \mbox{$<$ -173 dBc/Hz} for all offset frequencies higher than 600~Hz. This phase noise level is comparable to or below the lowest OFD microwave phase noise results yet achieved for offset frequencies greater than 100~Hz \cite{FortierOFD2013,Quinlan2013,syrte2016}.

The phase noise may be integrated to obtain an rms radian figure of merit. Integration from 1 Hz to 1 MHz yields \mbox{$\sim$0.35 rad$_{\mathrm{rms}}$} for the optical carrier, corresponding to 200~attoseconds of timing jitter. Further integration out to the optical Nyquist frequency of a shot-noise-limited floor assuming 1 mW of laser power (\mbox{-160 dBrad$^2$/Hz}) would only increase the integrated jitter to 210~attoseconds, integrated from 1~Hz to 140~THz. This should be compared with an estimate of the theoretical minimum for a thermal noise limited cavity with 1 mW output power. In this case the phase noise is \mbox{-9 dBrad$^2$/Hz} at 1 Hz and decreases as 1/f$^3$ until meeting a shot noise floor of \mbox{-160 dBrad$^2$/Hz}, yielding $\sim$160~attoseconds. Despite the demonstrated laser phase noise deviating from the shot noise and cavity thermal noise, $\sim$80 \% of its jitter may be attributed to these fundamental limits. This is because a large fraction of the jitter is due to the thermal noise from 1~Hz to 10~Hz, as can be seen in Fig. 2(c).

The phase noise spectrum has all the necessary information to calculate the Allan Deviation via integration with the appropriate kernel for each averaging period \cite{Stein1985}. This allows us to compare our cavity performance to the more common figure-of-merit for ultra-stable optical cavities. We find that our cavity reaches \mbox{$\sim2\times10^{-15}$} Allan Deviation at 0.1~s of averaging. For long term averaging the Allan Deviation is dominated by the drift due to uncompensated cavity temperature changes. To elucidate the different contributions to the Allan Deviation we performed an integration using the entire phase noise spectrum, shown in black in Fig. 3 and one using only the frequency band between 0.5~Hz and 50~kHz, shown in the blue curve. Note that the instability due to the additional noise in the 100~Hz to 1~kHz band becomes evident once the high-offset phase noise has been filtered out. Also, removing frequencies below 0.5~Hz partially compensates for the long-term drift. These results are plotted in Fig. 3, along with the calculated thermal noise limit at $1.6\times10^{-15}$.

\begin{figure}[htbp]
\centering
\includegraphics[width=3.3 in]{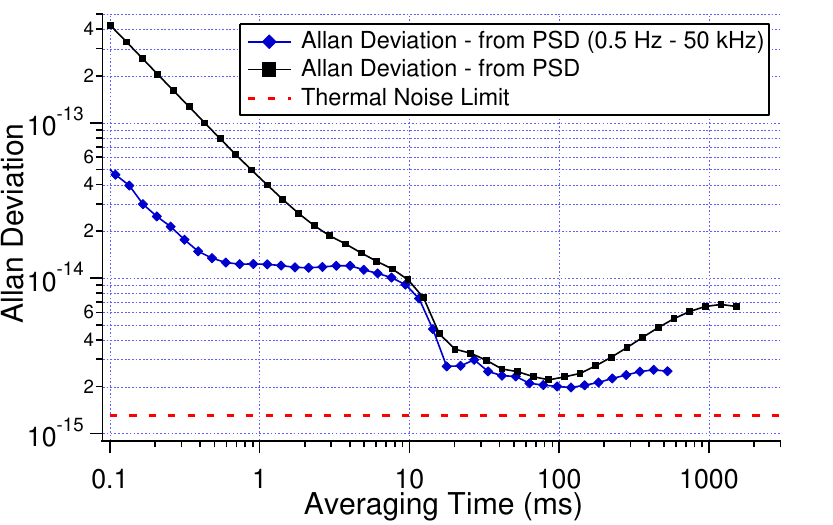}
\caption{Frequency stability derived from the phase noise. The stability reaches $2\times10^{-15}$, about 1.3 times the thermal noise limit.}
\label{fig:false-color}
\end{figure}

In conclusion, we have demonstrated a compact, thermal-noise limited, cavity-stabilized continuous wave laser which supports ultralow noise microwave generation. Our laser remains near thermal noise limited from 1~Hz to 1~kHz and can support 10~GHz microwave generation with phase noise below \mbox{-173 dBc/Hz} for all offset frequencies \mbox{$>600$~Hz}. Further improvement of the close-to-carrier noise may be accomplished with the use of crystalline mirror coatings \cite{Cole2013}, whereas a laser with lower free-running noise, such as a Brillouin laser \cite{Loh2015}, or self-injection locked semiconductor laser \cite{LiangMaleki2015}, should improve the noise far from carrier. With these improvements, a 25~mm long cavity capable of supporting 10 GHz phase noise approaching \mbox{-106 dBc/Hz} at 1~Hz and remaining below \mbox{-180 dBc/Hz} far from the carrier appears possible. Further work on minimization of the vibration sensitivity combined with straightforward long term temperature stabilization and residual amplitude modulation stabilization would improve the long-term stability making this cavity relevant for transportable optical atomic clock systems.

\textbf{Funding.} This work was funded by the DARPA PULSE Program and NIST. H. Leopardi is supported by an NDSEG Fellowship.

\textbf{Acknowledgments.} The authors wish to thank J. Sherman for support with the beat-note sampling as well as advice with the vacuum system, D. Hume, S. Cook and D. Leibrandt for providing the second reference laser at 1070 nm, D. Leibrandt for the loan of the rotatable optical breadboard, and M. Schioppo for assistance with finesse measurements and other discussions. We also would like to thank D. Leibrandt and W. Zhang for helpful comments on the manuscript. This work is a contribution of an agency of the US government and not subject to copyright in the USA.

\bibliographystyle{IEEE}
\bibliography{SmallCavityReferences}

\begin{thebibliography}{10}
\bibitem{NISTDisclaimer}
Product names are given for information purposes only and do not represent an
  endorsement by NIST.
\end{thebibliography}

\end{document}